\documentclass[twocolumn,english,showpacs,preprintnumbers,amsmath,amssymb,aps,prb,superscriptaddress,english]{revtex4}
\usepackage[T1]{fontenc}
\usepackage[latin9]{inputenc}
\usepackage{xcolor}
\usepackage{textcomp}
\usepackage{amsmath}
\usepackage{graphicx}
\usepackage{amssymb}
\PassOptionsToPackage{normalem}{ulem}
\usepackage{ulem}

\makeatletter

\providecommand{\tabularnewline}{\\}
\providecolor{lyxadded}{rgb}{0,0,1}
\providecolor{lyxdeleted}{rgb}{1,0,0}

\@ifundefined{textcolor}{}
{%
 \definecolor{BLACK}{gray}{0}
 \definecolor{WHITE}{gray}{1}
 \definecolor{RED}{rgb}{1,0,0}
 \definecolor{GREEN}{rgb}{0,1,0}
 \definecolor{BLUE}{rgb}{0,0,1}
 \definecolor{CYAN}{cmyk}{1,0,0,0}
 \definecolor{MAGENTA}{cmyk}{0,1,0,0}
 \definecolor{YELLOW}{cmyk}{0,0,1,0}
 }

\usepackage{dcolumn}
\usepackage{bm}
\makeatother

\usepackage{babel}
\usepackage{textcomp}

\definecolor{green}{rgb}{0.0,0.5,0.0}

\let\oldAA=\AA
\renewcommand{\AA}{\ensuremath{\mbox{\oldAA}}}

\makeatother

\usepackage{babel}

\begin{document}

\title{Bond disorder and breakdown of ballistic heat transport in the spin-1/2
antiferromagnetic Heisenberg chain as seen in Ca-doped SrCuO$_{2}$}

\author{N. Hlubek}

\thanks{These authors contributed equally to this work. }

\author{P. Ribeiro}

\thanks{These authors contributed equally to this work. }

\affiliation{IFW-Dresden, P.O. Box 270116, D-01171 Dresden, Germany}

\author{R. Saint-Martin}

\affiliation{Laboratoire de Physico-Chimie de L'Etat Solide, ICMMO, UMR8182, Universit\'{e}
Paris-Sud, F-91405 Orsay, France}

\author{S. Nishimoto}

\affiliation{IFW-Dresden, P.O. Box 270116, D-01171 Dresden, Germany}

\author{A. Revcolevschi}

\affiliation{Laboratoire de Physico-Chimie de L'Etat Solide, ICMMO, UMR8182, Universit\'{e}
Paris-Sud, 91405 Orsay, France}

\author{S.-L. Drechsler}

\affiliation{IFW-Dresden, P.O. Box 270116, D-01171 Dresden, Germany}

\author{G. Behr}

\thanks{Deceased.}

\author{J. Trinckauf}

\author{J. E. Hamann-Borrero}

\author{J. Geck}

\author{B. B\"{u}chner}

\author{C. Hess}

\email{c.hess@ifw-dresden.de}

\affiliation{IFW-Dresden, P.O. Box 270116, D-01171 Dresden, Germany}

\date{\today}

\pacs{75.40.Gb, 66.70.-f, 75.10.Kt, 75.10.Pq}
\begin{abstract}
We study the impact of a weak bond disorder on the spinon heat transport in the
$S=1/2$ antiferromagnetic (AFM) Heisenberg chain material $\rm Sr_{1-x}Ca_xCuO_{2}$.
We observe a drastic suppression in the magnetic heat conductivity $\kappa_{\mathrm{mag}}$ even at tiny disorder levels (i.e., Ca-doping levels), in stark contrast to previous findings for $\kappa_{\mathrm{mag}}$ of $S=1/2$ two-dimensional square lattice and two-leg spin-ladder systems, where a similar bond disorder has no effect on $\kappa_{\mathrm{mag}}$.
Hence, our results underpin the exceptional role of integrability of the $S=1/2$ AFM Heisenberg chain model and suggest that the bond disorder effectively destroys the ballistic nature of its heat transport. 
We further show that the suppression of $\kappa_{\mathrm{mag}}$ is captured by an effective spinon-impurity
scattering length, which exhibits the same doping dependence as the
long-distance exponential decay length of the spin-spin correlation as determined
by density-matrix renormalization group calculations.
\end{abstract}

\maketitle
\section{Introduction}
The transport properties of the integrable one-dimensional (1D) $S=1/2$ antiferromagnetic
(AFM) \textit{XXZ} chain model are attracting considerable attention because anomalous spin and heat transports have been predicted.\cite{Zotos1996,Zotos1997,Zotos1999,Kluemper2002,Heidrich2003,Heidrich-Meisner2007a,Sirker2009,Grossjohann2010,Znidaric2011,Mierzejewski2011,Prosen2011,Sirker2011}
Rigorous predictions concern the \textit{heat} transport of the model in the isotropic (Heisenberg) case. It is known to be \emph{ballistic} as the
consequence of integrability and fundamental conservation laws.\cite{Zotos1997,Zotos1999,Kluemper2002,Heidrich2003}
This means, a \textit{divergent} heat conductivity is expected. However,
in experimental realizations of this model, the observable spinon heat
conductivity $\kappa_{\mathrm{mag}}$ always has to be finite 
since extrinsic scattering processes due to defects and phonons are
inherent to all materials and mask the intrinsic behavior of a spin
chain. Nevertheless, a very large $\kappa_{\mathrm{mag}}$ has been
observed in a number of cuprates that realize $S=1/2$ spin chains.\cite{Sologubenko2001,Ribeiro2005,Hess2007a,Hess2007b,Kawamata2008,Hlubek2010a}
Especially, the material $\mathrm{SrCuO_{2}}$ is considered an excellent
realization of the 1D-AFM $S=1/2$ Heisenberg model (HM).\cite{Motoyama1996,Matsuda1997a,Zaliznyak2004}
For this material, we recently reported quasi ballistic spinon heat
transport with mean-free paths $l_0>1$\,\textmu{}m
for samples of extraordinary purity.\cite{Hlubek2010a}

Recently, several theoretical studies focused on the consequences of disorder for transport in 1D systems with controversial results.\cite{Damle2000,Laflorencie2004,Basko2006,Znidaric2008,Karahalios2009} Experimental studies are scarce and, so far, only concern heat transport studies on the 1D-AFM $S=1/2$ HM with site disorder (diagonal disorder), which cuts the chain into finite segments and, thus, rather trivially leads accordingly to a suppression of transport.\cite{Kawamata2008}
Here, we study the more subtle \textit{bond disorder} (off-diagonal disorder), which leaves the chains in the material intact and which recently has been investigated in nuclear magnetic resonance (NMR) experiments.\cite{Shiroka2011,Hammerath2011}
We induce this subtle type of disorder in SrCuO$_{2}$ by systematically substituting
isovalent Ca for Sr in tiny amounts. This off-chain lattice doping by small Ca$^{2+}$ ions induces a local lattice distortion which must result in a modulation, i.e., disorder, of the magnetic exchange constant $J$. In contrast to other doping schemes where the dopants occupy Cu sites within the magnetic structures and, thus, generate site disorder,\cite{Hess2003a,Hess2006,Kawamata2008} for this bond disorder, one generally expects only subtle changes in the magnetic transport properties of the system. In fact, previous studies show that the magnetic heat transport of the $S=1/2$ AFM square lattice\cite{Hess2003a} and the two-leg spin ladder systems,\cite{Hess2001} whose underlying spin models are non integrable, is unaffected by such kinds of disorder.
Radically different from these findings, in Ca-doped SrCuO$_{2}$, we observe a severe suppression
of $\kappa_{\mathrm{mag}}$ already at tiny doping levels ($\sim1$\%). Hence, these data
suggest that the disorder-induced
departure from integrability efficiently destroys the ballistic nature
of heat transport in the 1D-AFM $S=1/2$ HM. In the framework of a kinetic model, we
show that this suppression is
captured by an effective spinon-impurity term in the spinon mean-free
path, which surprisingly exhibits the same doping dependence as the
long-distance decay length of the spin-spin correlation as determined
by density-matrix renormalization group (DMRG) calculations. 

\section{Experimental Details}

We have grown centimeter sized single crystals of Sr$_{1-x}$Ca$_{x}$CuO$_{2}$
with $x=0$, 0.0125, 0.025, 0.05, 0.1 using the traveling-solvent
floating zone method.\citep{Revcolevschi1999} As starting materials
we used CuO (99.99\% purity), SrCO$_{3}$ (99\% purity) and CaCO$_{3}$
(99\% purity). Additionally, crystals with x=0, 0.0125 were grown
with all starting powders of 4N (99.99\%) purity. The crystallinity
and the doping profile were checked under polarized light and by energy dispersive 
x-ray spectroscopy.
A structural refinement was performed for samples of SrCuO$_{2}$ and Sr$_{0.9}$Ca$_{0.1}$CuO$_{2}$
in a single-crystal diffractometer. For the doped material, this yielded
a reduction in the cell volume by 1\% with respect to the undoped
compound in agreement with literature values.\citep{Ohashi1998}
Furthermore, we found a change in the Cu-O-Cu bond angle and the bond
distance of roughly 0.3\% 
\footnote{We find a bond length and angle of 3.910~$\mathrm{\AA}$ (3.896~$\mathrm{\AA}$)
and 175.5\textdegree{} (174.9\textdegree{}), respectively, for $x=0$
($x=0.1$).}. Since these data represent averages over the entire crystal volume,
we estimated the \textit{local} Ca-induced variation in the lattice
from density-functional calculations using the code \textsc{Quantum Espresso}.\cite{QE-2009} A $2\times2\times2$ supercell of SrCuO$_{2}$ doped
with 10\% Ca was relaxed with minimal symmetry assumptions. This yielded
a Cu-O-Cu bond length variation that is approximately the same as
the doping induced change in the mean Cu-O-Cu distances measured with
the single-crystal diffractometer. This consistency corroborates the
presence of a bond disorder in the material. 
For the transport measurements, rectangular samples with typical dimensions of $\left(3\cdot0.5\cdot0.5\right)\,\mbox{mm}^{3}$
for measurements of the heat conductivity along the principal axes
($\kappa_{a}$, $\kappa_{b}$, $\kappa_{c}$) were cut from the crystals
for each doping level with an abrasive slurry wire saw. The longest
dimension was taken parallel to the measurement axis. Measurements
of the thermal conductivity as a function of temperature $T$ in the
range of 7---300 K were performed with a standard four probe technique~\citep{Hess2003b}
using a differential Au/Fe-Chromel thermocouple for determining the
temperature gradient.

\begin{figure}

\includegraphics[width=8.5cm]{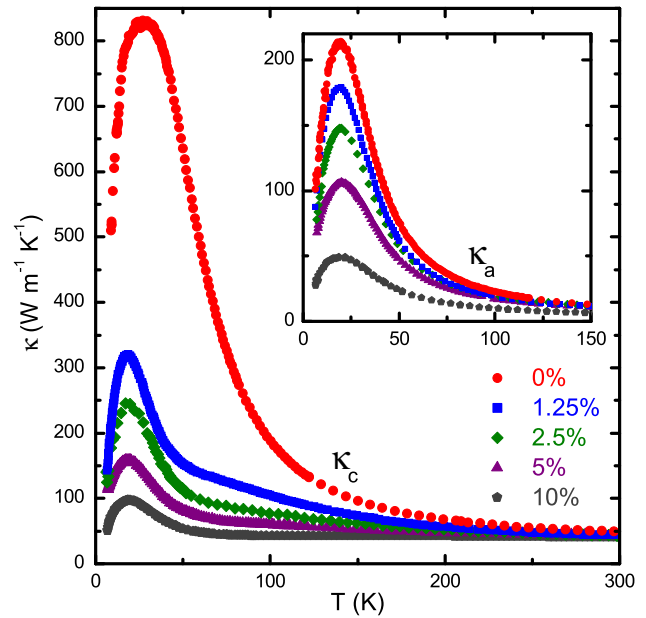}

\caption{(Color online) Thermal conductivity parallel to the spin chain ($\kappa_{c}$) for
Sr$_{1-x}$Ca$_{x}$CuO$_{2}$ at different doping levels. (Inset)
Thermal conductivity for the same doping levels perpendicular to the
spin chain ($\kappa_{a}$). \label{fig:kappa}}

\end{figure}

\section{Experimental data}
Figure~\ref{fig:kappa} shows our experimental results for $\kappa_{c}$
and $\kappa_{a}$ of Sr$_{1-x}$Ca$_{x}$CuO$_{2}$ for all doping
levels. In the pristine material ($x=0$), the heat conductivity parallel
to the spin chains $\kappa_{c}$ is strongly enhanced with respect
to the purely phononic heat conductivity perpendicular to the chains
$\kappa_{a}$. This is a result from quasi ballistic spinon heat transport
in the chains.\citep{Hlubek2010a} More specifically, the spinon
heat conductivity $\kappa_{\mathrm{mag}}$, which dominates the total
(spinon plus phonon) heat conductivity along the $c$ axis leads
to a broad peak at low temperatures ($\kappa_{c}\approx830\,\mathrm{Wm^{-1}K^{-1}}$
at $T\approx28$~K) followed by a strong decrease toward a still
significantly enhanced value ($\sim50\,\mathrm{Wm^{-1}K^{-1}}$) at
room temperature. 

Upon doping the material with Ca, the thermal conductivity perpendicular
to the spin chains ($\kappa_{a}$) is gradually suppressed. This is
the typical expectation of increased scattering by phonons off defects.~\citep{Berman}
Since the Ca$^{2+}$ impurities possess a smaller ionic radius and
mass as compared to Sr$^{2+}$, they act as defects for the phonons.%
\footnote{We also checked the thermal conductivity perpendicular to the chains
along $\kappa_{b}$. This gives very similar results apart from a
small anisotropy, which is already present in the undoped compound
and does not change in magnitude upon doping. %
} A much more dramatic change upon doping is observed in $\kappa_{c}$.
Already at the lowest doping level of 1.25\%~Ca, $\kappa_{c}$ is
suppressed strongly as compared to that of the pristine material, and
the overall curve shape of $\kappa_{c}(T)$ is changed completely.
The peak at low temperatures is now significantly smaller ($\kappa_{c}\approx320\,\mathrm{Wm^{-1}K^{-1}}$),
is much sharper, and is shifted toward lower temperatures ($T\approx18$\,K).
At about 50\,K, $\kappa_{c}$ shows a kink, above which, it decreases
much weaker and approaches $\kappa_{c}$ of the undoped compound for
even higher temperatures. Upon increasing the Ca doping, the absolute
value of the peak decreases, although its position stays constant.
The kink is gradually shifted to higher temperatures, and all of the
curves approach the curve of the undoped compound at room temperature.
This apparent saturation behavior is most visible at 10\%~Ca doping,
where a small maximum is followed by a practically constant thermal
conductivity. Nevertheless, the anisotropy between $\kappa_{a}$ and
$\kappa_{c}$ is still considerable since at 300\,K, $\kappa_{a}\approx4\,\mathrm{Wm^{-1}K^{-1}}$
and $\kappa_{c}\approx40\,\mathrm{Wm^{-1}K^{-1}}$. %
\footnote{Note, that measurements of $\kappa_{c}$ with 2N and 4N purity of
1.25\%~Ca doping did not show any significant difference. Therefore
the results for 2.5\%, 5\% and 10\%~Ca doping were obtained using
crystals of 2N purity.%
}

\section{Analysis}
The strong suppression of $\kappa_{c}$ implies that the Ca impurities
lead to a very strong suppression of the spinon heat conductivity
$\kappa_{\mathrm{mag}}$ already at very low doping levels ($x\approx0.01$)
because the doping scheme apparently only causes a moderate suppression
of the phononic heat conductivity, as seen in $\kappa_{a}$. Qualitatively,
this suggests that the doping-induced bond disorder leads to a significant
deviation in the spin system in Sr$_{1-x}$Ca$_{x}$CuO$_{2}$ from
the $S=1/2$ Heisenberg chain model, and, thus effectively destroys
the ballistic nature of the heat transport in the material. One might
conjecture that the connected departure from integrability can be
captured by an effective scattering process that describes the observed
suppression of $\kappa_{\mathrm{mag}}$ in our heat transport experiments.
In order to investigate this notion further, we analyze our data by
extracting $\kappa_{\mathrm{mag}}$ and calculating the spinon mean-free 
path $l_{\mathrm{mag}}$.

\begin{figure}[t]

\includegraphics[width=8.5cm]{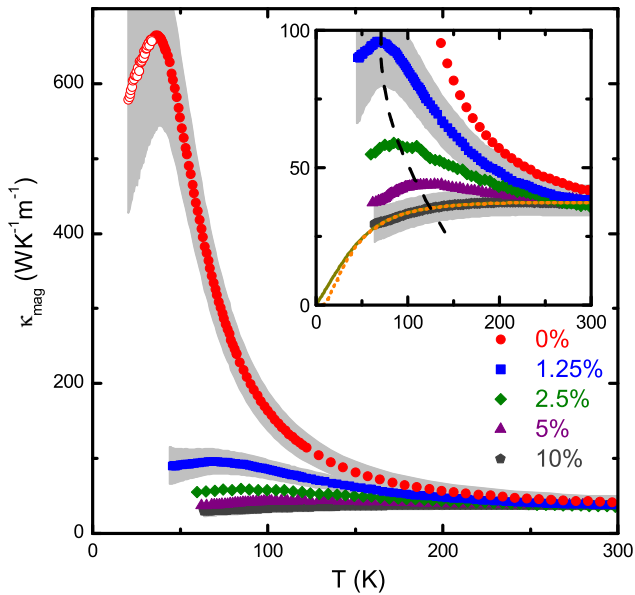}

\caption{(Color online) $\kappa_{\mathrm{mag}}$ for the different doping 
levels as a function of temperature. (Inset) Enlarged portion of the main 
plot shows the different doping levels in more detail. The black dashed line 
indicates the shift in the maximum to higher temperatures for higher dopings.
For 10\%~Ca doping, an extrapolation is shown without a spin gap (solid 
yellow line) and with a spin gap of 50\,K (dotted orange line). The shaded areas
show the uncertainty of the estimation of $\kappa_{\mathrm{mag}}$
due to the phononic background. All curves are shown only down to
a temperature for which the uncertainty of the estimation of $\kappa_{\mathrm{mag}}$
is still reasonably small. \label{fig:kappa mag}}

\end{figure}

\subsection{Spinon heat conductivity}
The first step of such an analysis~\citep{Hlubek2010a,Sologubenko2001}
consists in estimating the phononic part of $\kappa_{c}$ via $\kappa_{c,\mathrm{ph}}\approx\kappa_{a}$.
Then, the spinon thermal conductivity is given by $\kappa_{\mathrm{mag}}=\kappa_{c}-\kappa_{a}$~%
\footnote{Considering the small differences between $\kappa_{a}$ and $\kappa_{b}$
of $\approx15\%$ compared to the overall anisotropy of more than
a factor of 5 between $\kappa_{c}$ and $\kappa_{a}$, the validity
of this method seems to be justified.%
}. The doping-dependent evolution of $\kappa_{\mathrm{mag}}$ is shown
in Fig.~\ref{fig:kappa mag}. At low temperatures, $\kappa_{\mathrm{mag}}$
of the undoped compound shows a large maximum ($\kappa_{\mathrm{mag}}\approx665\,\mathrm{Wm^{-1}K^{-1}}$
at $T\approx36$\,K), which is followed by a steep decrease upon approaching
room temperature where the values of the curve nearly become constant
($\kappa_{\mathrm{mag}}\approx42\,\mathrm{Wm^{-1}K^{-1}}$). Doping
of 1.25\%~Ca leads to a severe suppression of $\kappa_{\mathrm{mag}}$
at low temperatures with a broadening of the maximum ($\sim96\,\mathrm{Wm^{-1}K^{-1}}$)
and a shift to 69\,K. Up to room temperature, the curve approaches
that of the undoped compound. Further increasing the doping continues
to decrease $\kappa_{\mathrm{mag}}$ at low temperatures and shifts
the increasingly broadened maximum to even higher temperatures. For
temperatures $T\gtrsim200$\,K, the values of $\kappa_{\mathrm{mag}}$
approach those of the undoped compound but remain smaller. %

\begin{figure}

\includegraphics[width=8.5cm]{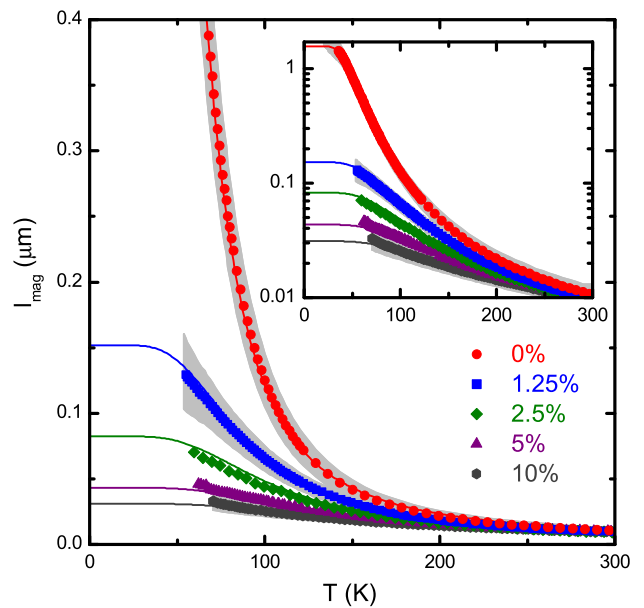} 

\caption{(Color online) $l_{\mathrm{mag}}$ of Sr$_{1-x}$Ca$_{x}$CuO$_{2}$ for different
levels of doping. Toward low temperatures, the error in the estimation
of $\kappa_{\mathrm{mag}}$ increases due to an increase in $\kappa_{\mathrm{ph}}$.
Thus, the values of $l_{\mathrm{mag}}$ are shown only at temperatures
above which the error in $l_{\mathrm{mag}}$ is reasonably small.
The solid lines were calculated according to Eq.~(\ref{eq:Umklapp}).
\label{fig:l mag}}

\end{figure}

\begin{table}
\begin{centering}
\begin{tabular}{c|c|c|c|c}
Ca content $x$ & $l_{0}$ ($\mathrm{\AA}$)  & $T_{u}^{*}$ (K)  & $A_{s}$ ($10^{-6}$K$^{-1}$m$^{-1}$) & $l_{0}$ ($c$)\tabularnewline
\hline 
0  & 15596  & 204  & 58.7 & 3989\tabularnewline
\hline 
0.0125  & 1519  & 204  & 65.5 & 389\tabularnewline
\hline 
0.025  & 824  & 204  & 66.3 & 211\tabularnewline
\hline 
0.05  & 433  & 204  & 54.8 & 112\tabularnewline
\hline 
0.1  & 311  & 204  & 50.3 & 80\tabularnewline
\end{tabular}
\par\end{centering}

\caption{Fit parameters for the mean-free paths according to Eq.~(\ref{eq:Umklapp}).
Additionally, $l_{0}$ is given in units of the lattice constant $c$.
\label{tab:Fit-parameters}}

\end{table}

\subsection{Spinon mean free path}
For each doping level, we use the derived $\kappa_{\mathrm{mag}}$
data to calculate the spinon mean-free path $l_{\mathrm{mag}}$ (see
Fig.~\ref{fig:l mag}) according to\citep{Sologubenko2001,Hess2007a,Kawamata2008,Hlubek2010a}
\begin{equation}
l_{\mathrm{mag}}(T)=\frac{3\hbar}{\pi N_{s}k_{B}^{2}T}\kappa_{\mathrm{mag}}\ \textcolor{blue}{,}\end{equation}
 with $N_{s}=4/ab$ as the number of spin chains per unit area. For all
doping levels, $l_{\mathrm{mag}}(T)$ decreases strongly with increasing
temperature which is the signature of spinon-phonon scattering.\citep{Hlubek2010a,Sologubenko2001}
However, upon increasing the doping level, $l_{\mathrm{mag}}$ exhibits
a systematic reduction, and the overall temperature-dependent change
at a given doping level becomes less pronounced, suggesting an increased
importance of spinon-defect scattering.\citep{Hlubek2010a} We test
this notion by applying Matthiesen's rule and decomposing $l_{\mathrm{mag}}$
into two respective contributions, i.e., $l_{\mathrm{mag}}^{-1}(T)=l_{0}^{-1}+l_{\mathrm{sp}}(T)^{-1}$.
Here, $l_{0}$ accounts for temperature-independent spinon-defect
scattering, whereas, $l_{\mathrm{sp}}(T)$ corresponds to $T$-dependent
spinon-phonon scattering. By using $l_{\mathrm{sp}}^{-1}\propto T\exp(-T_{u}^{*}/T)$
with a characteristic phonon energy scale $T_{u}^{*}$ on the order
of the Debye temperature,\citep{Sologubenko2001,Kawamata2008,Hlubek2010a}
we have \begin{equation}
l_{\mathrm{mag}}^{-1}(T)=l_{0}^{-1}+A_{s}T\exp\left(-T_{u}^{*}/T\right)\ ,\label{eq:Umklapp}\end{equation}
 with a proportionality constant $A_{s}$. Equation\ (\ref{eq:Umklapp})
is used to fit the $l_{\mathrm{mag}}$ data in Fig.~\ref{fig:l mag},
where $T_{u}^{*}$ first was determined for the pure compound and
then was used as a constant when fitting the data for finite-doping levels.
As can be seen in Fig.~\ref{fig:l mag}, the fits describe the data
very well (see Table~\ref{tab:Fit-parameters} for the fit parameters)
\footnote{Modeling the spinon scattering by thermally excited optical phonons\citep{Hlubek2010a,Hess2005} achieves a fit of similar quality and
comparable $l_{0}$ but with different $T^{*}$.%
}. A remarkable result is that $A_{s}$ also is constant nearly for
all doping levels with a root-mean-square deviation of around 10\%
of the arithmetic mean value. Thus, the effect of the bond disorder induced
by the Ca doping is described well by only one doping-dependent
parameter that describes the spinon-defect scattering $l_{0}$ and
one temperature-dependent spinon-phonon scattering mechanism, which
is independent of doping. 

Alternatively, for the afore discussed scattering, one might try to
attribute the suppressed $\kappa_{\mathrm{mag}}$ to a disorder-induced
depletion of thermally excited quasiparticles. In fact, recent NMR
measurements provided evidence for a spin gap $\Delta/k_{B}\approx50$~K
for 10\% Ca doping.\cite{Hammerath2011} However, the large $J$
makes the effect of this gap on the transport tiny. To illustrate
the negligible influence, the inset in Fig.~\ref{fig:kappa mag}
shows an estimate of the impact of a spin gap\cite{Hess2007a} on
$\kappa_{\mathrm{mag}}$ at 10\% Ca doping.

\begin{figure}
\includegraphics{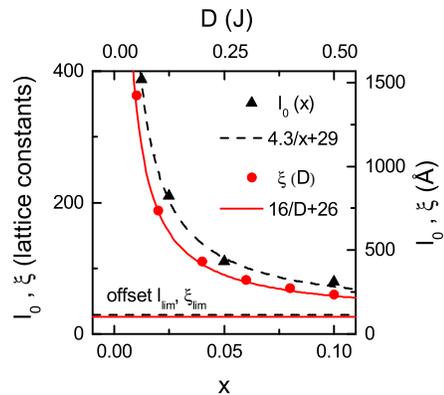}

\caption{(Color online) $l_{0}$ and $\xi$ as a function of doping $x$ (bottom axis) and
disorder parameter $D$ (top axis), respectively. The quantities are
fitted by $l_{0}=a_{1}/x+l_{\mathrm{lim}}$ ($a_{1}\approx4.3,\, l_{\mathrm{lim}}\approx29$)
and $\xi=a_{2}/D+\xi_{\text{lim}}$ ($a_{2}\approx16,\,\xi_{\mathrm{lim}}\approx26$).
It is worth noting that only a simple scaling factor for $D$ is necessary
to reach an almost perfect agreement between both curves. \label{fig:l0(x) vs. xi(D)}}

\end{figure}

Having established that the suppression of $\kappa_{\mathrm{mag}}$,
induced by the bond disorder, indeed can be described by an effective
spinon-defect scattering process, in Fig.~\ref{fig:l0(x) vs. xi(D)} we investigate
how the corresponding scattering length $l_{0}$ develops as a function
of doping. As can be inferred from the figure, $l_{0}$ decreases
systematically with doping, following $l_{0}=a_{1}/x+l_{\mathrm{lim}}$,
with the scattering strength $a_{1}$ and the offset $l_{\mathrm{lim}}$.
Note that, while $l_{0}\sim1/x$ is consistent with previous findings
for spin-defect doped chains, two-leg spin ladders and 2D square lattices,\citep{Kawamata2008,Hess2006,Hess2003a} the finding of a finite
offset $l_{\mathrm{lim}}$ is rather unexpected. Qualitatively, such
a finite offset means that the effect of the bond disorder, in terms
of defect like scattering, is strong at very small degrees of disorder
but becomes increasingly unimportant at larger degrees of disorder.
Hence, the effect of the bond disorder is significantly different
from that of strong site defects (diagonal disorder), which cut the
chains into finite segments.\citep{Kawamata2008} 

\subsection{Spin-spin correlation}
In order to obtain deeper insight into the effect of bond disorder
on the spin dynamics of the system, we investigated its influence
on the spin-spin correlation as a function of the distance~$z$.
The model Hamiltonian is written as \begin{eqnarray}
H=\sum_{i}(J+D\varepsilon_{i})\vec{S}_{i}\cdot\vec{S}_{i+1},\label{hamiltonian}\end{eqnarray}
 where $\vec{S}_{i}$ is a spin-$\frac{1}{2}$ operator at site $i$,
$J$ is the nearest-neighbor exchange interaction without bond disorder,
$\varepsilon_{i}$ is defined by a box probability distribution ${\cal P}(\varepsilon_{i})=\theta(1/2-|\varepsilon_{i}|)$
with the step function $\theta(x)$, and the disorder strength is controlled
by $D$. The spin-spin correlation functions $\left\langle \vec{S}_{i}\cdot\vec{S}_{i+z}\right\rangle $
are investigated using the DMRG.\citep{White1992} 
We study chains with $1024$ sites keeping $m=2000$
density-matrix eigenstates. Here, the open-boundary conditions are
used, and the distance $z$ is centered at the middle of the system.
We calculate the correlation functions by randomly sampling $300$
realizations of ${\cal P}(\varepsilon_{i})=\theta(1/2-|\varepsilon_{i}|)$
and taking an average of the results for each $D$ and $z$. For $D=0$,
the spin-spin correlation follows a $1/z$ law, which is expected
in the ballistic case. For $D>0$ the long-distance part can
be fitted just by $\exp\left(-z/\xi\right)$, with the free parameter
$\xi$. Since the commutator of the heat current operator and the
spin chain Hamiltonian is no longer zero for $D>0$, ballistic heat
transport no longer is expected. If one interprets the deviation
in the spin-spin correlation from the algebraic decay ($1/z$) as
a measure for the departure of the system from the clean one (i.e.
from integrability), the argument of the fitted exponential may fix
the crossover, with $z<\xi$ being ballistic and $z>\xi$ being diffusive.
Plotting $\xi$ as a function of $D$ (cf. Fig.~\ref{fig:l0(x) vs. xi(D)})
strikingly reveals that $\xi$ depends in the same manner on $D$
as $l_{0}$ does on $x$. Note, in particular, that the calculated
offset quantitatively almost perfectly matches that of the experiment,
i.e., $\xi\approx l_{0}$ in the strong disorder or large doping limit.
Furthermore, one may apparently assume $D\propto x$. Hence, in the
frame of this simple model, the long-distance decay length $\xi$
of the spin-spin correlation can be interpreted as a limit for the
effective spinon-defect scattering length $l_{0}$.

\section{Conclusion}
In conclusion, the doping-induced bond disorder in Sr$_{1-x}$Ca$_{x}$CuO$_{2}$
causes a severe suppression of the spinon heat conductivity which
is consistent with a disorder-induced departure of the underlying
spin model from integrability and, thus, the destruction of ballistic
heat transport. This interpretation is corroborated by the fact that the 
observed suppression apparently is a characteristic feature of this 
1D-AFM $S=1/2$ HM because no effect of bond disorder has been observed 
for a non integrable $S=1/2$ AFM square lattice and two-leg ladder systems. 
Using a simple kinetic model, we have shown that the suppression of 
$\kappa_{\mathrm{mag}}$ is described well by an effective spinon-defect 
scattering length, which can be related to the long-distance decay length 
of the spin-spin correlation as calculated by the DMRG method.

\begin{acknowledgments}
We thank W. Brenig, A. L. Chernyshev, and F. Heidrich-Meisner for fruitful
discussions. This work was supported by the Deutsche Forschungsgemeinschaft
through Grants No. HE3439/7, No. DRE269/3, and No. GE 1647/2-1, through the Forschergruppe
FOR912 (Grant No. HE3439/8), and by the European Commission through the
projects NOVMAG (Project No. FP6-032980) and the LOTHERM (Project No. PITN-GA-2009-238475). 
\end{acknowledgments}


\end{document}